\documentclass[twocolumn,showpacs,amsmath,amssymb]{revtex4}
\usepackage[dvips]{graphics}

\def\re#1{Re(#1)}
\def\im#1{Im(#1)}

\begin{document}
\title{Instability of $D$-dimensional extremally charged Reissner-Nordstr\o m(-de Sitter) black holes: Extrapolation to arbitrary $D$}
\author{R. A. Konoplya}\email{konoplya_roma@yahoo.com}
\affiliation{DAMTP, Centre for Mathematical Sciences, University of Cambridge, Wilberforce Road, Cambridge CB3 0WA, UK.}
\author{A. Zhidenko}\email{olexandr.zhydenko@ufabc.edu.br}
\affiliation{Centro de Matem\'atica, Computa\c{c}\~ao e Cogni\c{c}\~ao, Universidade Federal do ABC (UFABC),\\ Rua Aboli\c{c}\~ao, CEP: 09210-180, Santo Andr\'e, SP, Brazil}

\begin{abstract}
In our earlier work (PRL 103 (2009) 161101) it was shown that nonextremal highly charged Reissner-Nordstr\o m-de Sitter black holes are gravitationally unstable in $D > 6$-dimensional space-times. Here, we find accurate threshold values of the $\Lambda$-term at which the instability of the extremally charged black holes starts. The larger $D$ is, the smaller is the threshold value of $\Lambda$.
We have shown that the ratio $\rho = r_{h}/r_{cos}$ (where $r_{cos}$ and $r_{h}$ are the cosmological and event horizons) is proportional to $e^{-(D-4)/2}$ at the onset of instability for $D=7, 8, \ldots11$, implying that the same law should fulfill for arbitrary $D$. This is numerical evidence that extremally charged Reissner-Nordstr\o m-de Sitter black holes are gravitationally unstable for $D>6$, while asymptotically flat extremally charged Reissner-Nordstr\o m black holes are stable for all $D$. The instability is not connected to the horizon instability discussed recently in the literature, and, unlike the later one, develops also outside the event horizon, that is, it can be seen by an external observer. In addition, for the nonextremal case through fitting of the numerical data we obtained an approximate analytical formula which relates values of charge and the $\Lambda$-term at the onset of instability.
\end{abstract}
\maketitle

\section{Introduction.}

Stability of a metric against small space-time perturbations, usually termed as \emph{classical}, \emph{gravitational} or \emph{dynamical} stability (in order to distinguish from the thermodynamical stability, which has quantum origin), is a basic criteria of viability of a black hole model.
While stability of four-dimensional black hole solutions has been very well (but not exhaustively) investigated, stability of higher-dimensional black holes has been less studied. In higher than four dimensions, absence of the traditional uniqueness theorem allows for a wide class of solutions with event horizons of various topologies \cite{Emparan:2008eg}. This makes the problem of stability even more appealing, because it allows one to discard inviable solutions. In addition, perturbation, quasinormal modes and stability of higher-dimensional asymptotically de Sitter (dS) and anti-de Sitter (AdS) black holes and branes play an important role in the dS/CFT \cite{Strominger:2001pn} and AdS/CFT correspondences \cite{Son:2007vk}.

Linear stability of higher-dimensional generalization of Schwarzschild black holes (given by the Tangherlini solution) was analytically proven for arbitrary $D$ by Kodama and Ishibashi \cite{Ishibashi:2003ap}. For a broad class of D-dimensional black holes allowing for the electric charge and lambda term the perturbation equations were reduced to the wavelike form \cite{Kodama:2003kk}. However, the analytical treatment of the stability problem looks
intractable in $D>4$ if either the charge or lambda term is not zero. Therefore, in a number of cases the stability problem was attacked numerically through the investigation of dominant resonance frequencies of the perturbation (quasinormal modes) in frequency domain, and through modeling of the evolution of perturbation in the time domain.

In this way, i.e., by numerical investigation of a black-hole quasinormal spectrum, stability of various of higher-dimensional black holes was analyzed.
In \cite{Konoplya:2007jv} and \cite{Konoplya:2008rq} numerical evidence of linear stability of $D=5, 6, \ldots11$-dimensional Schwarzschild-AdS, Reissner-Nordstr\o m-AdS and Schwarzschild-de Sitter black holes was found in the Einstein-Maxwell theory.
In our earlier paper \cite{Konoplya:2008au} it was shown that $D>6$-dimensional nonextremal Reissner-Nordstr\o m-de Sitter black holes are unstable when the electric charge and $\Lambda$-term are large enough. Some other cases, including higher curvature corrections, dilaton, etc were considered in the literature (reviewed in \cite{Konoplya:2011qq} and \cite{Ishibashi:2011ws}).
Yet, as the above works used numerical treatment of the problem, each fixed value of $D$ was treated ``one by one'' leaving aside the general proof of (in)stability for arbitrary $D$.

Here we shall study the (in)stability of extremally charged Reissner-Nordstr\o m(-de Sitter) black holes numerically, and generalize the results to an arbitrary number of space-time dimensions $D$. In \cite{Konoplya:2008au} the instability was demonstrated for $D=7, 8, \ldots11$ highly charged but \emph{nonextremal} black holes. Here we shall find critical values of the cosmological constant $\Lambda$ which correspond to the threshold of instability of the \emph{extremally charged} Reissner-Nordstr\o m-de Sitter solution for $D=7, 8, \ldots11$. Then, we shall extrapolate the data to higher $D$ and find a remarkably simple dependence of critical $\Lambda$ on $D$, also indicating that the asymptotically flat extremal Reissner-Nordstr\o m black holes must be linearly stable for arbitrary $D$. Using the numerical data from \cite{Konoplya:2008au} for the nonextremally charged case here we shall show that the threshold values of black-hole parameters can be fitted by a simple analytical formula as well.

The paper is organized as follows: Sec.~II gives basic formulae for the space-time metric, the master perturbation equations, and the time-domain integration method which we used. Sec.~III is devoted to the analysis of numerical data and its extrapolation to small values of the cosmological terms and large $D$. Sec.~IV describes the procedure of fitting of numerical data for the threshold parameters by an analytical formula. Sec.~V, the conclusion, is a brief summary of results on stability of static higher-dimensional black holes in the Einstein-Maxwell theory studied by now.

\section{Basic formulae and the time-domain integration}

The metric of a $D$-dimensional Reissner-Nordstr\o m-de-Sitter black hole has the form
\begin{equation}\label{metric}
ds^2=f(r)dt^2-\frac{dr^2}{f(r)}-r^2(d\theta^2+sin^2\theta
d\phi^2),
\end{equation}
where
\begin{equation}\label{metric-function}
f(r)=1-\frac{2M}{r^{D-3}}+\frac{Q^2}{r^{2D-6}}-\frac{2\Lambda r^2}{(D-2)(D-1)}.
\end{equation}

The perturbation equation can be treated separately for scalars, vectors, and tensors and can be reduced to the wavelike form \cite{Kodama:2003kk} for each case,
\begin{equation}\label{hyperbolic}
\left(\frac{\partial^2}{\partial t^2}-\frac{\partial^2}{\partial r_*^2}\right)\Psi(t,r)=-V(r)\Psi(t,r),
\end{equation}
where \emph{the tortoise coordinate} $r_*$ is defined as
\begin{equation}\label{tortoise}
dr_*=\frac{dr}{f(r)}.
\end{equation}

We shall be interested here in the scalar type of gravitational perturbations, for which the effective potential has a negative gap, and, thereby, the instability is possible. This type of perturbations is referred to as the ``-'' type (see (5.61) of \cite{Kodama:2003kk}).
The effective potential $V(r)$ depends on the following parameters: the number of space-time dimensions $D$, black hole mass $M$, electric charge $Q$, lambda-term $\Lambda$, multipole number $\ell=2,3,4\ldots$. Usually the instability occurs at the lowest multipole number $\ell=2$. In our case higher multipoles are more stable. Therefore, here we shall show numerical data only for $\ell=2$ perturbations of the ``-'' type.

We shall measure all the quantities in the units of the horizon radius ($r_h=1$). Thus we define
\begin{equation}\label{horizon-units}
2M=1+Q^2-2\Lambda/(D-2)(D-1).
\end{equation}

In order to parameterize the cosmological constant and the black-hole charge we introduce two dimensionless parameters,
$$0\leq\rho=r_h/r_{cos}<1\quad\mbox{and}\quad 0\leq\sigma=r_i/r_h\leq1,$$ where $r_{cos}$ and $r_i$ are, respectively, the cosmological and inner horizons.

Then, using the following equations,
$$f(\sigma)=f(1/\rho)=0,$$
we can find $\Lambda$ and $Q$ in terms of these parameters
\begin{eqnarray}\label{paramaterization}
\Lambda&=&\frac{\rho^2}{2}\frac{(D-2)(D-1)(1-\sigma^{D-3}\rho^{D-3})}{\displaystyle\left(\frac{1-\rho^{D-1}}{1-\rho^{D-3}}\right)-\sigma^{D-3}\rho^{D-1}\left(\frac{1-\sigma^{D-1}}{1-\sigma^{D-3}}\right)},\\\nonumber
Q^2&=&\sigma^{D-3}\frac{\displaystyle\left(\frac{1-\rho^2}{1-\rho^{D-1}}\right)-\sigma^{D-3}\left(\frac{1-\sigma^2}{1-\sigma^{D-1}}\right)}{\displaystyle\left(\frac{1-\sigma^{D-3}}{1-\sigma^{D-1}}\right)-\sigma^{D-3}\rho^{D-1}\left(\frac{1-\rho^{D-3}}{1-\rho^{D-1}}\right)}.
\end{eqnarray}
It is easy to see from (\ref{paramaterization}) that, when $\Lambda$ or $Q$ is zero, the corresponding dimensionless parameter, $\rho$ or $\sigma$, vanishes. By definition, $\sigma=1$ corresponds to the extremal charge of a black hole. As $\rho\rightarrow1$ the size of a black hole approaches to its extremal value in the de Sitter universe. Values of the parameters at the onset of instability we shall call \emph{critical} or \emph{threshold} values.

We shall study the evolution of the black hole perturbations of scalar ``-'' type ($\ell=2$) in time domain using a numerical characteristic integration method \cite{Gundlach:1993tp}, that uses the light-cone variables $u = t - r_\star$ and $v = t + r_\star$. In the characteristic initial value problem, initial data are specified on the two null surfaces $u = u_{0}$ and $v = v_{0}$. The discretization scheme we used is
\begin{eqnarray}\label{d-uv-eq}
\Psi(N) &=& \Psi(W) + \Psi(E) - \Psi(S) -\\\nonumber&&-\Delta^2\frac{V(W)\Psi(W) + V(E)\Psi(E)}{8} + \mathcal{O}(\Delta^4) \ ,
\end{eqnarray}
where we have the following definitions for the points: $N =(u + \Delta, v + \Delta)$, $W = (u + \Delta, v)$, $E = (u, v + \Delta)$ and $S = (u,v)$. This method was tested for finding accurate values of the damped quasinormal modes
(see for instance \cite{dampedQNMs} and references therein). Recently it has been also adopted for finding \emph{unstable}, growing, quasinormal modes in \cite{Konoplya:2008yy} for black strings and for Gauss-Bonnet black holes \cite{Konoplya:2008ix2}. The agreement between the time-domain and accurate Frobenius methods (as well as with the WKB method in the region of its validity \cite{WKB}) is excellent. To test the reliability of the method, we increased the precision of the whole numerical procedure and decreased the grid of integration: unchanging of the obtained profiles of $\Psi$ signifies that we have reached sufficient accuracy of computation.

\section{Numerical data: extrapolations to small $\rho$ and to  large $D$}

\begin{figure*}
\resizebox{\linewidth}{!}{\includegraphics*{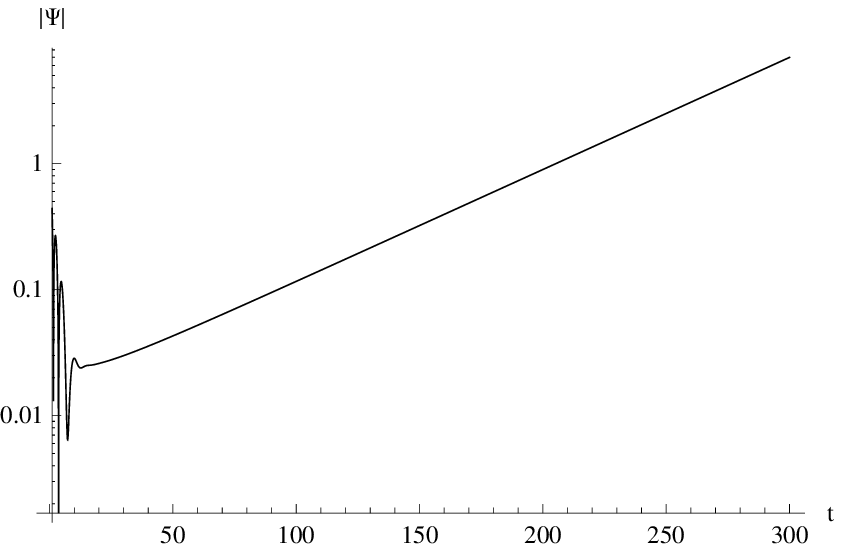}\includegraphics*{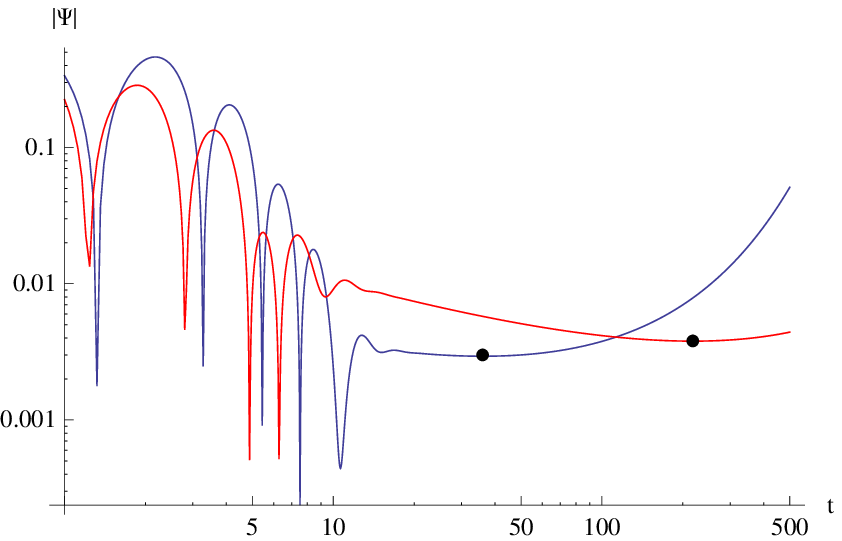}}
  \caption{Left panel (logarithmic scale): Typical unbounded growth of the signal for $\rho>\rho_c$ ($D=9$, $\ell=2$, $\rho=0.4$). Right panel (log-log scale): Two profiles for $D=9$, which correspond to $\rho=0.3$ (blue) and $\rho=0.2$ (red); the points show the moment after which the unbounded growth is observed. The smaller the value of $\rho$ is (i.e. the closer to the threshold value $\rho_c$), the later the moment at which the signal starts growing.}
\label{1}
\end{figure*}

\begin{figure*}
\resizebox{\linewidth}{!}{\includegraphics*{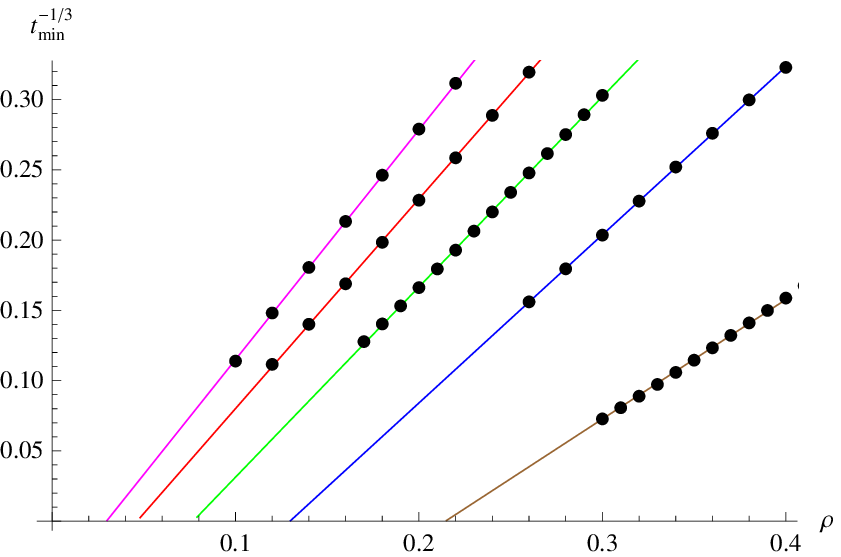}\includegraphics*{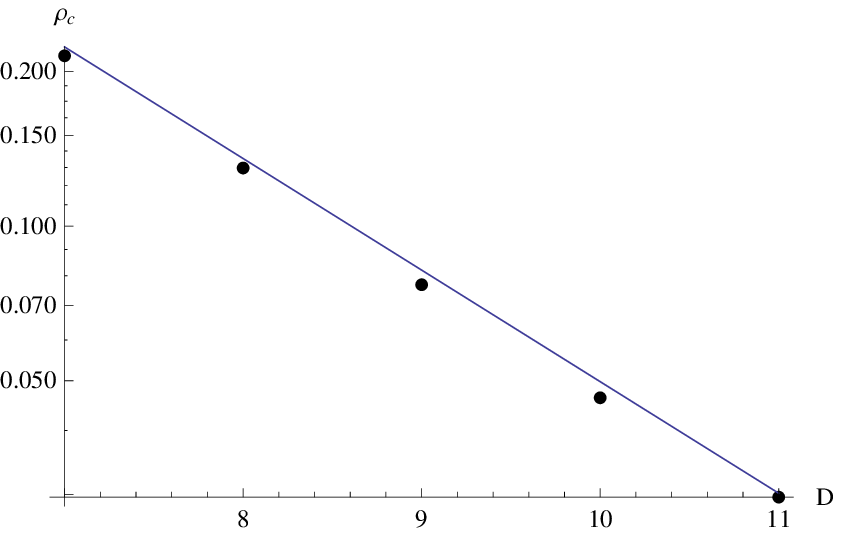}}
\caption{Left panel: $t_{min}^{-1/3}$ as a function of $\rho$ for (from right to left) $D=7$ (brown), $D=8$ (blue), $D=9$ (green), $D=10$ (red), and $D=11$ (magenta), and $t_{min}$ is the moment at which we observe the beginning of the unbounded growth of the signal. Solid lines were obtained by fitting by linear functions of our numerical data (black points). The threshold points of stability $\rho_c$, which corresponds to the intersections of the lines with the horizontal axes, are presented for each $D$ in the right panel using the logarithmic scale together with our fit for large $D$ (solid line) $\rho_c=e^{-(D-4)/2}$.}
\label{2}
\end{figure*}

A thorough study of profiles of $|\Psi|$ at sufficiently late time in the region of instability shows that the moment at which the growth of the signal starts depends on the value of $\rho$. At a fixed spatial coordinate the smaller $\rho$, the later is the moment at which growth of the profile begins (Fig.~\ref{1}). With a personal computer and reasonable computation time, one cannot find the asymptotic tails for time, which is much later than $t \sim 10^3 r_h$. Thus, an extrapolation to large $t$ (and small $\rho$) is necessary. Fortunately, all the numerical values obtained here for the ``starting moments'' of instability (let us designate them as $t_{min}$) obey the inverse cubic law very well:
\begin{equation}
(\rho - \rho_c)\propto t_{min}^{-1/3},
\end{equation}
where the constant $\rho_c$ depends on $D$  and is positive for $D =7, 8,\ldots11$. Thus, numerically found $\rho_c$ are \emph{threshold values of $\rho$ corresponding to the onset of instability for the extremally charged black hole}. Then, it is natural to try to find these critical values of $\rho$ for any $D$, and the extrapolation to higher $D$ also obeys a remarkably simple law (see Fig.~\ref{2}):

\begin{equation}\label{rho-c}
\rho_c \approx A e^{- \frac{D-4}{2}},
\end{equation}
where the constant $A$ is close to unity. By fitting the numerical values of $\rho_c$ found for $D=7,8,9,10,11$ we obtain $A \approx 0.96$.

Thus, we conclude that the critical value of the normalized cosmological constant $\rho$ corresponding to the onset of instability goes to zero as the number of space-time dimensions $D$ goes to infinity. That is, the \emph{extremally charged Reissner-Nordstr\o m-de Sitter black holes unstable for all $D> 6$, while the pure Reissner-Nordstr\o m solution must be stable for arbitrary $D$}.
The latter statement may look too strong and simply means that if there is an instability for the extremal Reissner-Nordstr\o m solution at some large $D$, it has a different nature from the one we observed for Reissner-Nordstr\o m-de Sitter black holes. If, contrary to our observation, the constant $\rho_c$ were negative at some $D$, that would mean that one should expect an instability of the pure Reissner-Nordstr\o m solution ($\rho=0$) at this value of $D$.

We computed time-domain profiles at a value of spatial coordinate $r$ fixed somewhere near the peak of the potential barrier. Although the values of $t_{min}$ almost do not depend on the choice of $r$, we considered profiles of $|\Psi|$ at the same fixed $r$ for all values of $\rho$ on Figs.~\ref{1} and~\ref{2}. However, the choice of $r$ at the potential peak allows one to reduce the period of initial outburst to minimum.

We take a constant as initial conditions on the null surface from the side of the black hole and a Gaussian at the other null surface. We also computed time-domain profiles with another choice of initial conditions, namely, a constant at both null surfaces. Although the profiles and, consequently, $t_{min}$ depend softly on the choice of initial conditions, we observe the same behavior of $t_{min}$ as a function of $\rho$. By fitting these points we find the same critical values of $\rho$. We conclude, therefore, that the onset of instability does not depend on the choice of the initial conditions, and the point $r$ where the time-domain profile is considered. It is possible that some fine-tuning of the initial conditions could lead to unbounded growth of the signal just after the initial outburst for all $\rho>\rho_c$. However, for the considered initial conditions the growing mode appears with a relatively small amplitude. That is why we observe growing at relatively late time, i.e., when the stable (damped) modes are faded out. As $\rho$ approaches $\rho_c$ the growing rate of the unstable mode decreases and, therefore, it dominates at later time. That is why $t_{min}$ increases as $\rho\rightarrow\rho_c$.

The instability of extremal Reissner-Nordstr\o m and Kerr black holes found recently in \cite{Aretakis:2011gz,Lucietti:2012xr,Lucietti:2012sf} is qualitatively different from the one we observe here.
The instability discussed in \cite{Aretakis:2011gz,Lucietti:2012xr,Lucietti:2012sf} propagates only along the event horizon and, therefore, at the linear order cannot be observed by an external observer, so that for the observer the system still should look stable \cite{Konoplya:2013rxa}. The instability which we analyzed here propagates outside the event horizon and thereby may serve as criteria for viability of the black hole model or as an indication of bifurcation towards a new solution.

\section{Extrapolation to arbitrary $D$ for nonextremal charges}

A simple formula (\ref{rho-c}) for critical values of $\rho$ motivated us to look at the nonextremal case as well and study critical values of $\sigma$ for fixed $\rho$ as a function of $D$. Taking data from our earlier paper \cite{Konoplya:2008au}, we have observed that for each $\rho>\rho_c$ the critical value of $\sigma$ can be approximated by a simple function
\begin{equation}\label{fit-sigma-D}
1-\sigma_c\approx B-Ce^{-D},
\end{equation}
where constants $B$ and $C$ depend only on $\rho$. According to numerical data, $C\gg B$ for each $\rho$, so that for $D\leq6$ the right-hand side is always negative: i.e., $\sigma_c>1$ for $D\leq6$. That is why we prefer the following equivalent exponential fitting:
\begin{equation}\label{fit-sigma-D-alt}
1-\sigma_c\approx B-\widetilde{C}e^{6-D}.
\end{equation}
Comparing $B$ and $\widetilde{C}$ for various values of $\rho$, we have observed a remarkable relation:
$$\widetilde{C}\approx 2B\propto \rho^3.$$

\begin{figure}
\resizebox{\linewidth}{!}{\includegraphics*{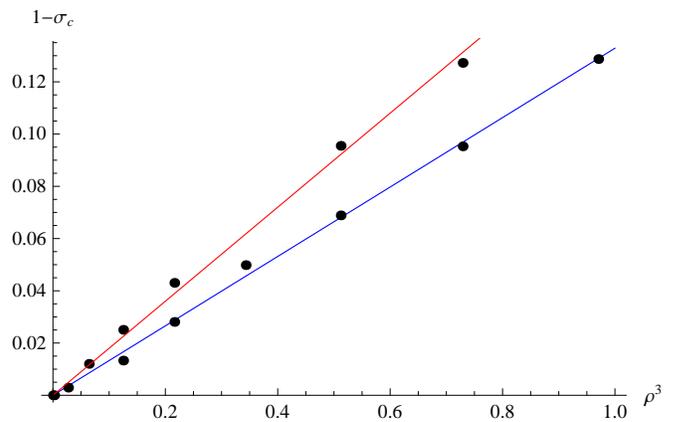}}
  \caption{Critical values of $\sigma$ as functions of $\rho^3$ and linear fits for $D=8$ (blue lower line) and $D=10$ (red upper line).}
\label{3}
\end{figure}

By fitting the critical value of $\sigma_c$ (see Fig.~\ref{3}) as
$$1-\sigma_c\approx\alpha_D\rho^3,$$
we found the following coefficients:
\begin{center}
\begin{tabular}{r|l}
$D$&$\alpha_D$\\
\hline
$7$&$0.05$\\
$8$&$0.133$\\
$9$&$0.168$\\
$10$&$0.180$\\
$11$&$0.188$\\
\end{tabular}
\end{center}

\begin{figure}
\resizebox{\linewidth}{!}{\includegraphics*{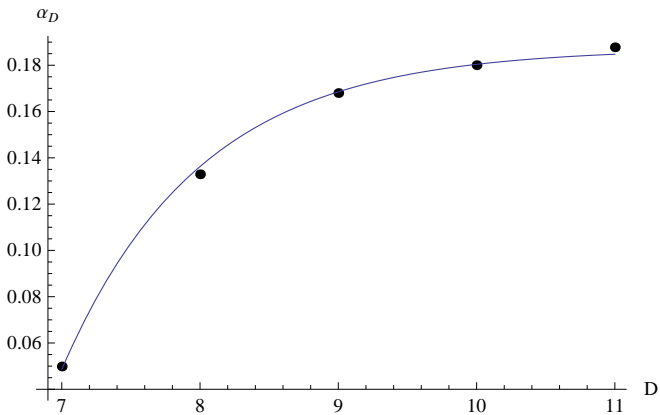}}
  \caption{Values of $\alpha_D$ and the fit in the following form: $\alpha_D\approx 0.187-0.377e^{6-D}$.}
\label{4}
\end{figure}

The obtained values of $\alpha_D$ (see Fig.~\ref{4}) can be well fitted by the following relation:
$$\alpha_D\approx0.187-0.377e^{6-D}.$$
Within the numerical accuracy of \cite{Konoplya:2008au} the value of $\sigma_c$ can be given in the following form
\begin{equation}\label{sigma-c}
\sigma_c\approx1-\alpha_D\rho^3\approx1-\left(\frac{4\rho}{7}\right)^3\left(1-2e^{6-D}\right).
\end{equation}
Although (\ref{sigma-c}) has an elegant form and explains the stability of the Reissner-Nordstr\o m-de Sitter black holes for $D\leq6$, let us note that it is an approximation obtained by fitting of the numerical data of \cite{Konoplya:2008au} by a simple combination of elementary functions. Thus, even the $\sigma_c=1$ limit as $\rho\rightarrow\rho_c$ cannot be reproduced within (\ref{sigma-c}). Apparently subdominant terms, which may play an essential role when $\rho\rightarrow\rho_c$, are lacking here. Nevertheless, we believe that this simple formula may be useful for comparison of our numerically found region of instability with further, more precise studies.

\begin{figure*}
\resizebox{\linewidth}{!}{\includegraphics*{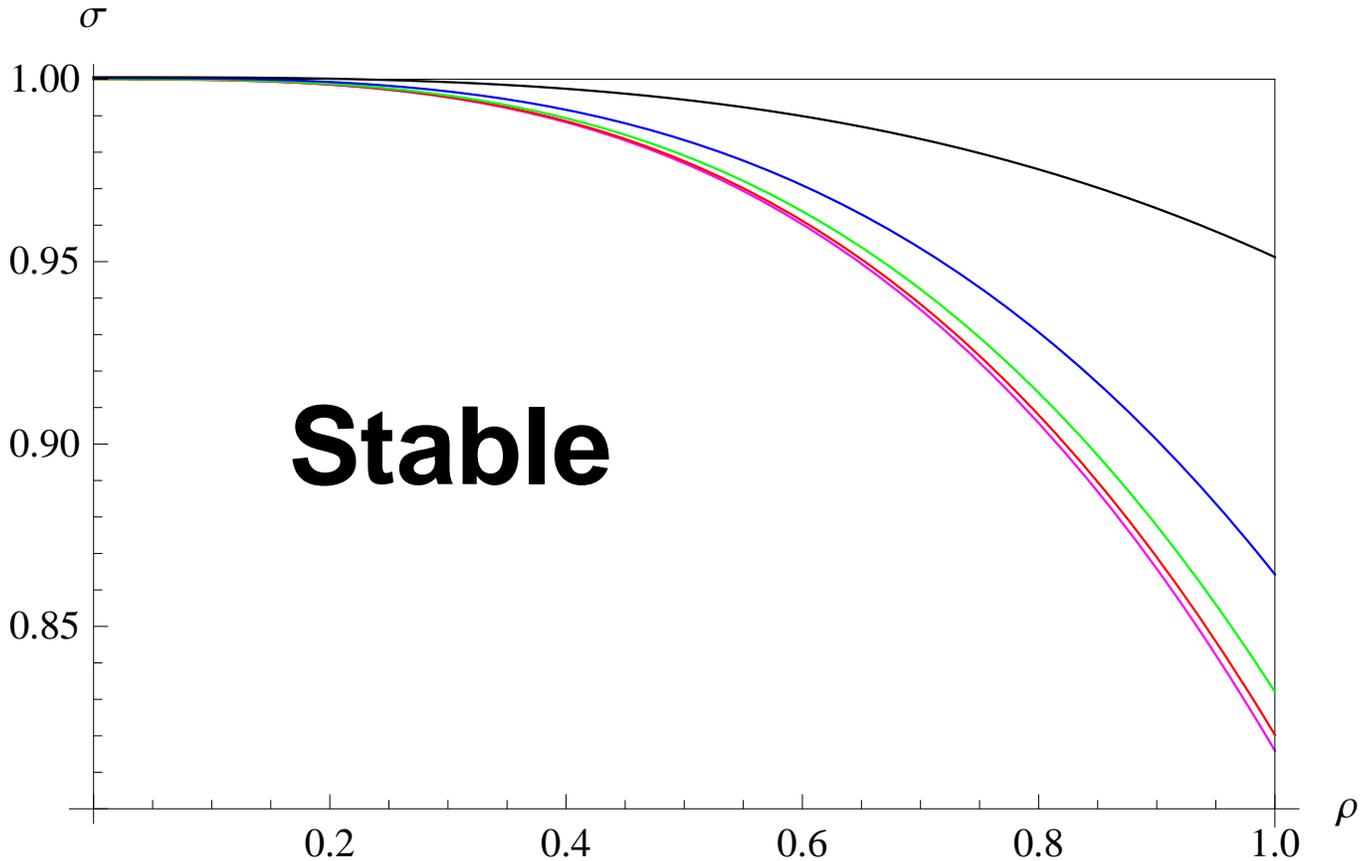}}
  \caption{Parametric region of stability of Reissner-Nordstr\o m-de Sitter black holes in the $\rho$-$\sigma$ plane for $D=7$ (black upper line), $D=8$ (blue line), $D=9$ (green line), $D=10$ (red line), and $D=11$ (magenta lower line).}
\label{5}
\end{figure*}

For $\rho\rightarrow1$, we obtain the asymptotic values for $\sigma_c$, which are lower than those found in \cite{Cardoso:2010rz}, where only a sufficient condition of instability was considered.
Finally, taking into account that $\rho_c^3$ is small for $D>6$, in order to plot our estimation for the region of instability (see Fig.~\ref{5}), we use the following formula, which provides the correct value of $\sigma_c$ for $\rho=\rho_c$:
\begin{eqnarray}\label{sigma-c-ref}
\sigma_c=1-\left(\left(\frac{4\rho}{7}\right)^3-\left(\frac{4\rho_c}{7}\right)^3\right)\left(1-2e^{6-D}\right),\\\nonumber
\rho_c=e^{-\frac{D-4}{2}}\leq\rho<1.
\end{eqnarray}

The obtained analytical formula has been deduced by fitting of the numerical data. An analytical approach of \cite{Emparan:2013moa} could possibly be an alternative method to analyze of the quasinormal spectrum in the large $D$ limit.

\section{Conclusions}

\begin{figure}
\resizebox{\linewidth}{!}{\includegraphics*{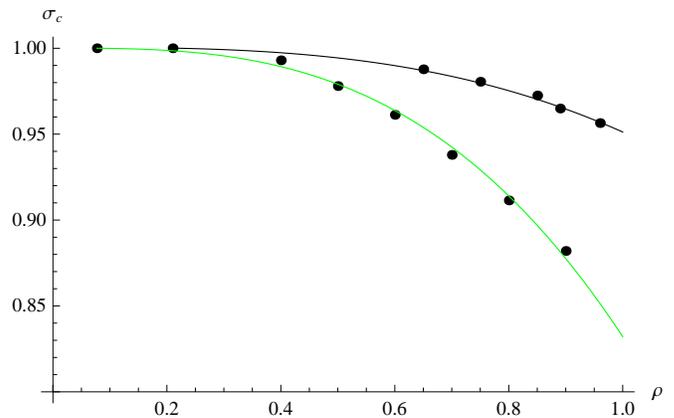}}
  \caption{Comparison of numerical values of $\sigma_c$ from \cite{Konoplya:2008au} (black points) and formula (\ref{sigma-c-ref}) for $D=7$ (black upper line) and $D=9$ (green lower line).}
\label{6}
\end{figure}

Here we showed that the instability of higher-dimensional extremally charged black holes in the de Sitter world obey a couple of remarkable laws:
\begin{itemize}
\item  For smaller values of the cosmological constant, the instability starts at later times, and the beginning of the growth of $|\Psi|$ obeys a simple law $(\rho-\rho_c) \propto t_{min}^{-1/3}$, where $\rho_c$ is the critical value of $\rho$ corresponding to the onset of instability and $t_{min}$ is the moment at which the signal starts growing.

\item The critical $\rho$ at the onset of instability is proportional to  $e^{-\frac{D-4}{2}}$  for $D>6$. This indicates the instability of all $D>6$ RN-dS black holes (when $\rho>\rho_c$) and stability of the pure RN solution.

\end{itemize}

The found relations for the threshold parameters are valid for various choices of initial conditions in the time-domain integration, in the same manner as it takes place for quasinormal modes.

\begin{table}
\caption{Stability of static black holes in the Einstein-Maxwell gravity.}\label{tabl-I}
\begin{tabular}{|c|c|c|c|}
  \hline
  charge & $\Lambda=0$ & $\Lambda>0$ & $\Lambda<0$ \\
  \hline
  S ($Q=0$) & stable & stable & stable \\
  \hline
  RN ($Q \neq 0$)& stable\footnote{The extremal RN has a kind of instability which is limited by the event horizon \cite{Aretakis:2011gz}. The physical meaning of this instability for the external observer apparently could be studied in the nonlinear approximation.} & unstable for $D>6$ & stable\footnote{The RN-AdS black holes, being stable in the Einstein-Maxwell theory, have a region of instability in the supergravity \cite{Gubser-Mitra}.} \\
  \hline
\end{tabular}
\end{table}

The current status of (in)stability for static higher-dimensional black holes in the Einstein-Maxwell gravity is briefly summarized in table~\ref{tabl-I}.

In addition, we obtained the approximate analytical formula (\ref{sigma-c}) relating parameters of a nonextremally charged black hole at the onset of instability (see Fig.~\ref{6} for comparison of the formula with the numerical data).

At the onset of instability, the dominant unstable mode has $\re{\omega}=\im{\omega}=0$, which means that the perturbation approaches a constant at late times. The destiny of the unstable space-time can be learnt only within the fully nonlinear approach. The shape of a slightly perturbed black hole at the onset of instability was considered by us in \cite{Konoplya:2008au}.

\section*{Acknowledgments}
This work was supported by the European Commission grant through the Marie Curie International Incoming Program.
R. A. K. acknowledges support of his visit to Universidade Federal do ABC by FAPESP.
A.~Z. was supported by Conselho Nacional de Desenvolvimento Cient\'ifico e Tecnol\'ogico (CNPq).


\begin{thebibliography}{80}
\bibitem{Emparan:2008eg}
  R.~Emparan and H.~S.~Reall,
  Living Rev.\ Rel.\  {\bf 11}, 6 (2008)
  [arXiv:0801.3471 [hep-th]].
\bibitem{Strominger:2001pn}
  A.~Strominger,
  JHEP {\bf 0110}, 034 (2001)
  [hep-th/0106113].
\bibitem{Son:2007vk}
  D.~T.~Son and A.~O.~Starinets,
  Ann.\ Rev.\ Nucl.\ Part.\ Sci.\  {\bf 57}, 95 (2007)
  [arXiv:0704.0240 [hep-th]].
\bibitem{Ishibashi:2003ap}
  A.~Ishibashi and H.~Kodama,
  Prog.\ Theor.\ Phys.\  {\bf 110}, 901 (2003)
  [hep-th/0305185].
\bibitem{Kodama:2003kk}
  H.~Kodama and A.~Ishibashi,
  Prog.\ Theor.\ Phys.\  {\bf 111}, 29 (2004)
  [hep-th/0308128].
\bibitem{Konoplya:2007jv}
  R.~A.~Konoplya and A.~Zhidenko,
  Nucl.\ Phys.\ B {\bf 777}, 182 (2007)
  [hep-th/0703231 [HEP-TH]].
\bibitem{Konoplya:2008rq}
  R.~A.~Konoplya and A.~Zhidenko,
  Phys.\ Rev.\ D {\bf 78}, 104017 (2008)
  [arXiv:0809.2048 [hep-th]].
\bibitem{Konoplya:2008au}
  R.~A.~Konoplya and A.~Zhidenko,
  Phys.\ Rev.\ Lett.\  {\bf 103}, 161101 (2009)
  [arXiv:0809.2822 [hep-th]].
\bibitem{Konoplya:2011qq}
  R.~A.~Konoplya and A.~Zhidenko,
  Rev.\ Mod.\ Phys.\  {\bf 83}, 793 (2011)
  [arXiv:1102.4014 [gr-qc]].
\bibitem{Ishibashi:2011ws}
  A.~Ishibashi and H.~Kodama,
  Prog.\ Theor.\ Phys.\ Suppl.\  {\bf 189}, 165 (2011)
  [arXiv:1103.6148 [hep-th]].
\bibitem{Gundlach:1993tp}
  C.~Gundlach, R.~H.~Price, and J.~Pullin,
  Phys.\ Rev.\  D {\bf 49}, 883 (1994)
 [arXiv:gr-qc/9307009].
\bibitem{dampedQNMs}
  H.~Ishihara, M.~Kimura, R.~A.~Konoplya, K.~Murata, J.~Soda, and A.~Zhidenko,
  Phys.\ Rev.\  D {\bf 77}, 084019 (2008)
  [arXiv:0802.0655 [hep-th]].
\bibitem{Konoplya:2008yy}
  R.~A.~Konoplya, K.~Murata, J.~Soda and A.~Zhidenko,
  Phys.\ Rev.\ D {\bf 78}, 084012 (2008)
  [arXiv:0807.1897 [hep-th]].
\bibitem{Konoplya:2008ix2}
  R.~A.~Konoplya and A.~Zhidenko,
  Phys.\ Rev.\  D {\bf 77}, 104004 (2008)
  [arXiv:0802.0267 [hep-th]].
\bibitem{WKB}B.~F.~Schutz and C.~M.~Will, Astrophys.\ J.\ Lett. {\bf 291}, L33 (1985);
S.~Iyer and C.~M.~Will Phys.\ Rev.\  D {\bf 35}, 3621 (1987);
R. A. Konoplya, Phys.\ Rev\ D {\bf 68}, 024018 (2003); J.\ Phys.\ Stud.\ {\bf 8}, 93 (2004).
\bibitem{Aretakis:2011gz}
  S.~Aretakis,
  J.\ Funct.\ Anal.\  {\bf 263}, 2770 (2012)
  [arXiv:1110.2006 [gr-qc]];
  Commun.\ Math.\ Phys.\  {\bf 307}, 17 (2011)
  [arXiv:1110.2007 [gr-qc]].
\bibitem{Lucietti:2012xr}
  J.~Lucietti, K.~Murata, H.~S.~Reall, and N.~Tanahashi,
  JHEP {\bf 1303}, 035 (2013)
  [arXiv:1212.2557 [gr-qc]].
\bibitem{Lucietti:2012sf}
  J.~Lucietti and H.~S.~Reall,
  Phys.\ Rev.\ D {\bf 86}, 104030 (2012)
  [arXiv:1208.1437 [gr-qc]].
\bibitem{Konoplya:2013rxa}
  R.~A.~Konoplya and A.~Zhidenko,
  Phys.\ Rev.\ D {\bf 88}, 024054 (2013)
  [arXiv:1307.1812 [gr-qc]].
\bibitem{Cardoso:2010rz}
  V.~Cardoso, M.~Lemos, and M.~Marques,
  Phys.\ Rev.\ D {\bf 80}, 127502 (2009)
  [arXiv:1001.0019 [gr-qc]].
\bibitem{Emparan:2013moa}
  R.~Emparan, R.~Suzuki, and K.~Tanabe,
  JHEP {\bf 1306}, 009 (2013)
  [arXiv:1302.6382 [hep-th]].
\bibitem{Gubser-Mitra}
  S.~S.~Gubser and I.~Mitra,
  JHEP {\bf 0108}, 018 (2001)
  [hep-th/0011127].
\end{thebibliography}
\end{document}